\documentclass[12pt]{article}

\usepackage{scicite}

\usepackage{times}

\usepackage{graphicx} 

\usepackage{adjustbox,lipsum}
\usepackage{array, booktabs, makecell}
\usepackage[font=footnotesize]{caption}

\usepackage{float}

\newenvironment{sciabstract}{%
\begin{quote} \bf}
{\end{quote}}

\title{A Multiple Filter Based Neural Network Approach to the Extrapolation of Adsorption Energies on Metal Surfaces for Catalysis Applications}

\author
{Asif J. Chowdhury,$^{1}$ Wenqiang Yang,$^{2}$ Kareem E. Abdelfatah,$^{1}$ Mehdi Zare,$^{2}$\\Andreas Heyden,$^{2\ast}$ Gabriel A. Terejanu$^{3\ast}$\\
\\
\normalsize{$^{1}$Department of Computer Science and Engineering, University of South Carolina}\\
\normalsize{$^{2}$Department of Chemical Engineering, University of South Carolina}\\
\normalsize{$^{3}$Department of Computer Science, University of North Carolina Charlotte}\\
\\
\normalsize{$^\ast$To whom correspondence should be addressed; E-mail:  heyden@cec.sc.edu, gterejan@uncc.edu}
}

\date{}

\begin{document} 


\baselineskip24pt


\maketitle


\begin{sciabstract}
  Computational catalyst discovery involves the development of microkinetic reactor models based on estimated parameters determined from density functional theory (DFT). For complex surface chemistries, the cost of calculating the adsorption energies by DFT for a large number of reaction intermediates can become prohibitive. Here, we have identified appropriate descriptors and machine learning models that can be used to predict part of these adsorption energies given data on the rest of them. Our investigations also included the case when the species data used to train the predictive model is of different size relative to the species the model tries to predict - an extrapolation in the data space which is typically difficult with regular machine learning models. We have developed a neural network based predictive model that combines an established model with the concepts of a convolutional neural network that, when extrapolating, achieves significant improvement over the previous models.
\end{sciabstract}


\section*{Introduction}

Computational catalyst discovery typically requires the development of a microkinetic model based on parameters determined by density functional theory (DFT) calculations~\cite{structdep} of all reaction intermediates~\cite{compcatal}. To minimize the cost of calculating the energies for each reaction intermediate and transition state on different active site models, linear scaling relations~\cite{lscl2,lscl1,linearscale} have been proposed which use a few easily computable descriptors, such as the carbon atom adsorption energy, on different active site models to generate volcano curves on catalyst activity~\cite{theohet}. However, even the DFT computations for only the intermediate species on a number of surfaces require, for a large reaction network with many intermediates, a significant number of expensive calculations. In this work, our goal was to build a predictive framework that would train on the energies of some of the surface species and predict on the rest, which can significantly reduce the computational overhead when working with a complex microkinetic model with a large number of surface species. In addition to that, we also investigated appropriate predictive models for extrapolation of adsorption energies in terms of the size of the species, i.e, when the training data and the prediction set contain different sized molecules. This is typically challenging because machine learning models, while performing satisfactorily during interpolation (when training and testing set come from the same area of the feature space), do not work as well for extrapolation~\cite{difdom} (when training and testing set come from non-overlapping regions of the feature space). 

In this paper, we have worked with two data sets of adsorption energies, both containing reaction intermediates consisting of carbon, oxygen, and hydrogen atoms. One of them contains $247$ larger C4 species, i.e, molecules with at least four carbon atoms and variable numbers of oxygen and hydrogen, obtained from the hydrodeoxygenation of succinic acid on Pt(111). The other contains $29$ smaller C2 and C3 species, i.e, molecules made up of two or three carbon atoms along with some oxygen and/or hydrogen, obtained from a reaction network of decarboxylation and decarbonylation reactions of propionic acid on Pt(111)~\cite{hdoprop}. All calculations were done using the PBE-D3 functional. Two types of predictive analysis were performed - interpolation on the bigger C4 dataset, i.e, training on some of the C4 species and predicting on the rest of them; and extrapolation from the C4 data set to the C2 and C3 data set, i.e, training on the full C4 data and predicting the adsorption energies for the C2 and C3 species. While extrapolation to longer chain molecules is in principle most relevant, we do not possess a C5 dataset and the C2 and C3 datasets are too small to be used for extrapolation to C4 species. Nevertheless, extrapolation from C4 to C2 and C3 is technically as challenging as extrapolation to longer chain molecules and we expect all of our conclusions to also be valid when extrapolating to longer chain molecules.

Predicting properties of some chemical entity using machine learning~\cite{gennn,dftml} involves solving two related subproblems - discovery of effective features or descriptors, and using a proper machine learning model that, together with the chosen descriptor, works best for the specific task at hand~\cite{selfpap1}. Here, we are trying to predict the adsorption energies of surface intermediates and hence, the descriptors are essentially some form of molecular fingerprints. Many different kinds of fingerprints or fingerprint generation schemes have been proposed in previous studies - Coulomb matrix~\cite{cmatrix} and bag-of-bonds~\cite{bob} using distance measures between the atomic coordinates of the species; atom centered radial or angular symmetry functions~\cite{behlersymfun,behlerref2,behlerref3,norssubnet}; non-coordinate based fingerprints that take into account features of a molecule which can be extracted from the chemical formula or SMILES notation~\cite{extconfp,chemoinf,invmol,molbench}; generation of fingerprints from molecular graph structure~\cite{graphconv,convfp} where the atoms and the bonds are considered as the nodes and edges of a graph, respectively, and fingerprints corresponding to a target property learned using back-propagation etc. Fingerprints based on SMILES or graph have the desirable property over the coordinate based descriptors that the DFT or other semi-empirical methods need to be applied only on the training data; for the rest of the data for which the adsorption energies are unknown, their molecular notation is all that the predictive model would need to make the predictions. In contrast, the coordinate based methods would need reliable atomic coordinates even for the species on the prediction set which would require some form of expensive calculations - ones we wish to minimize in the first place. Most commonly used machine learning models have been kernel based models such as kernel ridge regression~\cite{krrml} and different neural network based models such as graph convolution~\cite{graphconv,convfp}, recurrent neural network~\cite{rnnmolfp}, 3D convolutional neural network which reads the 3D spatial coordinates of the molecule~\cite{3dcnnbio}, or additive atomic contribution through atomic subnetworks~\cite{gennn,behlersymfun}. 

In our investigations for interpolation of adsorption energies, we have studied both the coordinate based and SMILES based descriptors along with a variety of machine learning models. Our results indicate that simple molecular descriptors that capture the nearest neighbor information across the species from the SMILES notation, paired with kernel based models can perform as good as coordinate based descriptors such as Coulomb matrix or bag-of-bonds with a mean absolute error (MAE) of $0.14$ eV. However, for extrapolation, the choice of descriptor is more complex - descriptors based on pairwise or triplet atomic distances such as Coulomb matrix or bag-of-bonds have the disadvantage that they are not size extensible. For data with different sized species, smaller ones have to be padded with zeros that make the learning difficult. In this case, constant sized molecular fingerprints~\cite{lilienconstdesc} are more suitable. Our results, however, suggest that the predictive errors are still quite high for these descriptors. A different kind of approach, which is atom centered and where each atom's neighborhood information (its pairwise and triplet distances from other atoms) is treated as the atomic fingerprint and fed to a small neural network where these subnetworks from each atom share their weights, and then all of the atoms' contributions are added up to get the final energy, is size extensible. We found this method to work better than the other methods for extrapolation with MAEs of around $0.4$ eV. However, the error is still large and we have sought ways to improve upon this model. One improvement was to include SMILES based atomic fingerprints over the coordinate based ones, and the second contribution, that helped to get significantly smaller extrapolation errors, was to treat the small atomic subnetworks like a filter of a convolution neural network and use multiple of these filters. This method had extrapolation MAE of $0.23$ eV.

\section*{Methodology}

Molecular fingerprints used in our investigations can be categorized into three classes: first, coordinate based Coulomb matrix and bag-of-bonds that use pairwise distances between the atoms in the molecule to generate the fingerprint; second, flat fingerprints based on the number of different bond counts inside the molecule that can be read from the chemical formula or SMILES notation; third, atom centered fingerprints that are based on the local neighborhood around an atom which is calculated either by distance measures between the atomic coordinates or by the number of different bond types for the atom that can be read from the molecular notation. Machine learning models that we have used can also be divided into three categories: generalized linear models such as linear regression, ridge regression (which uses L2 regularizer), LASSO (which uses L1 regularizer); kernel based models such as kernel ridge regression (KRR), support vector regression (SVR), Gaussian processes (GP); and artificial neural networks (ANN).

\subsection*{Coulomb Matrix and Bag-of-bonds}
The Coulomb matrix (CM) method first creates a symmetric matrix where the off-diagonal element $C(i,j)$ is a function of the distance measures between the i-th and j-th atoms and also their atomic numbers. The diagonal elements are a function of the atomic number of the corresponding atoms. The sorted eigenvalues of the matrix forms the molecular fingerprint. The Bag-of-bond (BoB) method takes the off-diagonal lower triangle of the symmetric matrix formed in CM and puts the entries corresponding to each atom type pair in a bag, sorts the entries inside each bag and concatenates the bags to form the fingerprint vector. We have found these methods to typically work well for interpolative predictions among the same sized species. However, for data with variable sized species, one needs to pad the entries of the matrix for smaller species with zeros. This limits their usefulness for size-extrapolation predictions. Detailed methods for building CM and BoB are described in the supporting information.

\subsection*{Flat Molecular Fingerprint}
\begin{figure}[!thb]
\centering
   \includegraphics[scale=0.45]{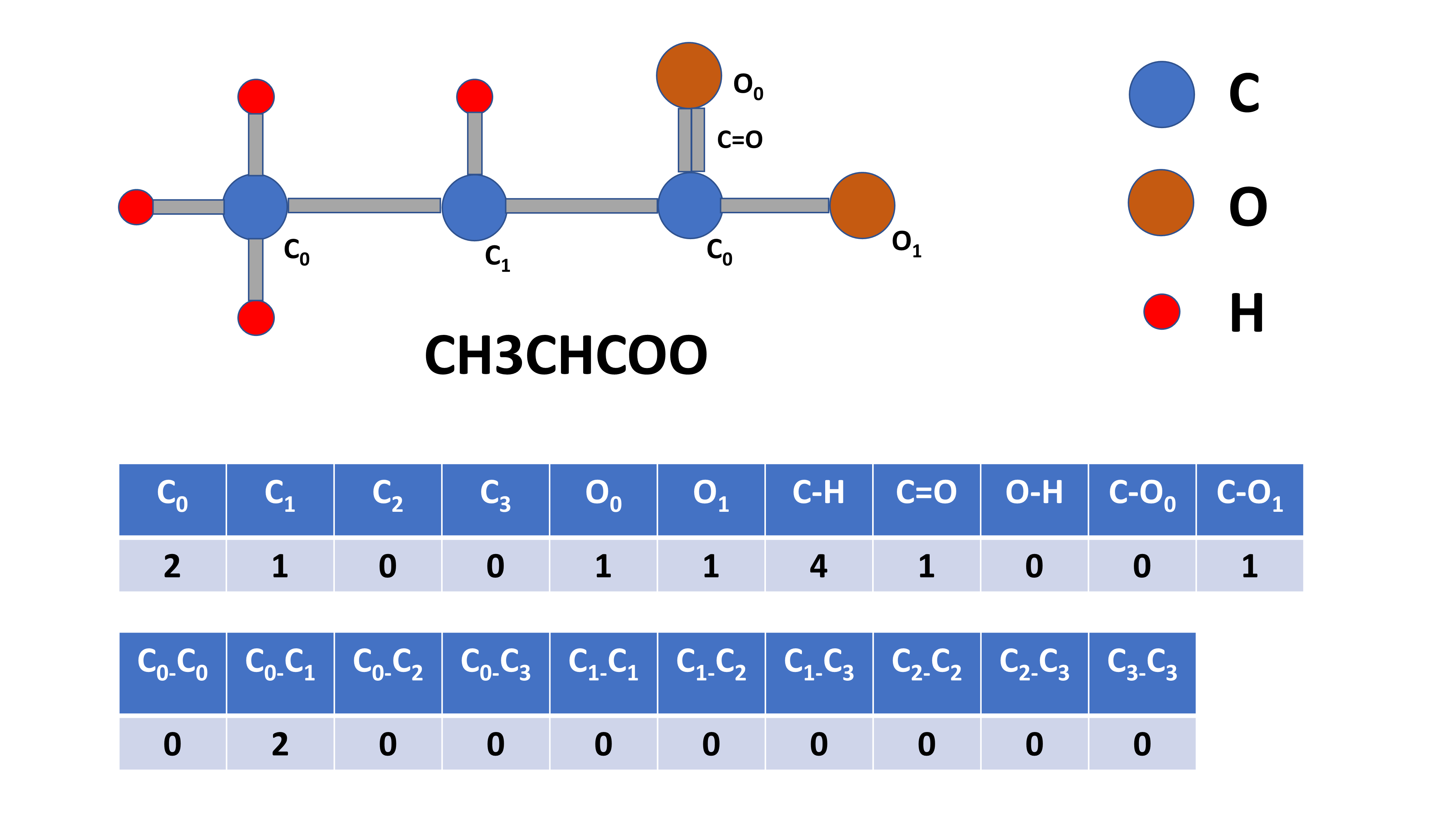}
\caption{Molecular fingerprint for the surface species $CH3CHCOO$. Here, $C_0$ denotes a saturated carbon (no free valence). $C_1$, $C_2$, and $C_3$ denote carbon atoms with one, two, and three free valencies, respectively. Similarly, $O_0$ is a saturated oxygen whereas $O_1$ is an oxygen atom with one free valence. The fingerprint vector (shown at the bottom of the image) contains the number of different saturated or unsaturated atoms, and the number of bonds between them.} \label{fig:molfp}
\end{figure}

Fingerprints generated from the SMILES notation of the adsorbed species encode the connectivity among the atoms inside the molecule. The encoding can capture the number of different types of atoms or bonds by looking into the nearest neighbors of each atom or upto some specified distance. In our study, we have built a simple scheme that looks into the nearest neighbors of the atoms and counts the number of different atom types an atom is connected to, and then accumulates the results in a fingerprint vector. The proposed fingerprint is shown in Figure~\ref{fig:molfp}). Here, atom types are divided into subclasses by the number of free valencies an atom has, e.g, instead of just looking at how many carbon-carbon bonds are present, the fingerprint captures how many saturated carbon atoms are single bonded to a carbon with one free valence, or how many oxygens are double bonded to a saturated carbon atom and so on. This is a constant sized molecular descriptor as the length of the vector remains the same for smaller or bigger molecules. As will be discussed later, this method works well for interpolation and works better than CM or BoB for extrapolation, but the extrapolation error was still quite large. Both, the flat molecular fingerprint and CM/BoB, can be fed to any regular ML model such as linear model, kernel based model or fully connected feed forward neural network. In each case, the ML model takes as input the molecular fingerprint vector and outputs the target real value (in our case, adsorption energy). We have also tried the extended connectivity based fingerprint (ECFP)~\cite{extconfp} which produces fixed length vectors from the SMILES notations of the molecules.

\begin{figure}[H]
\centering
	\includegraphics[scale=0.58]{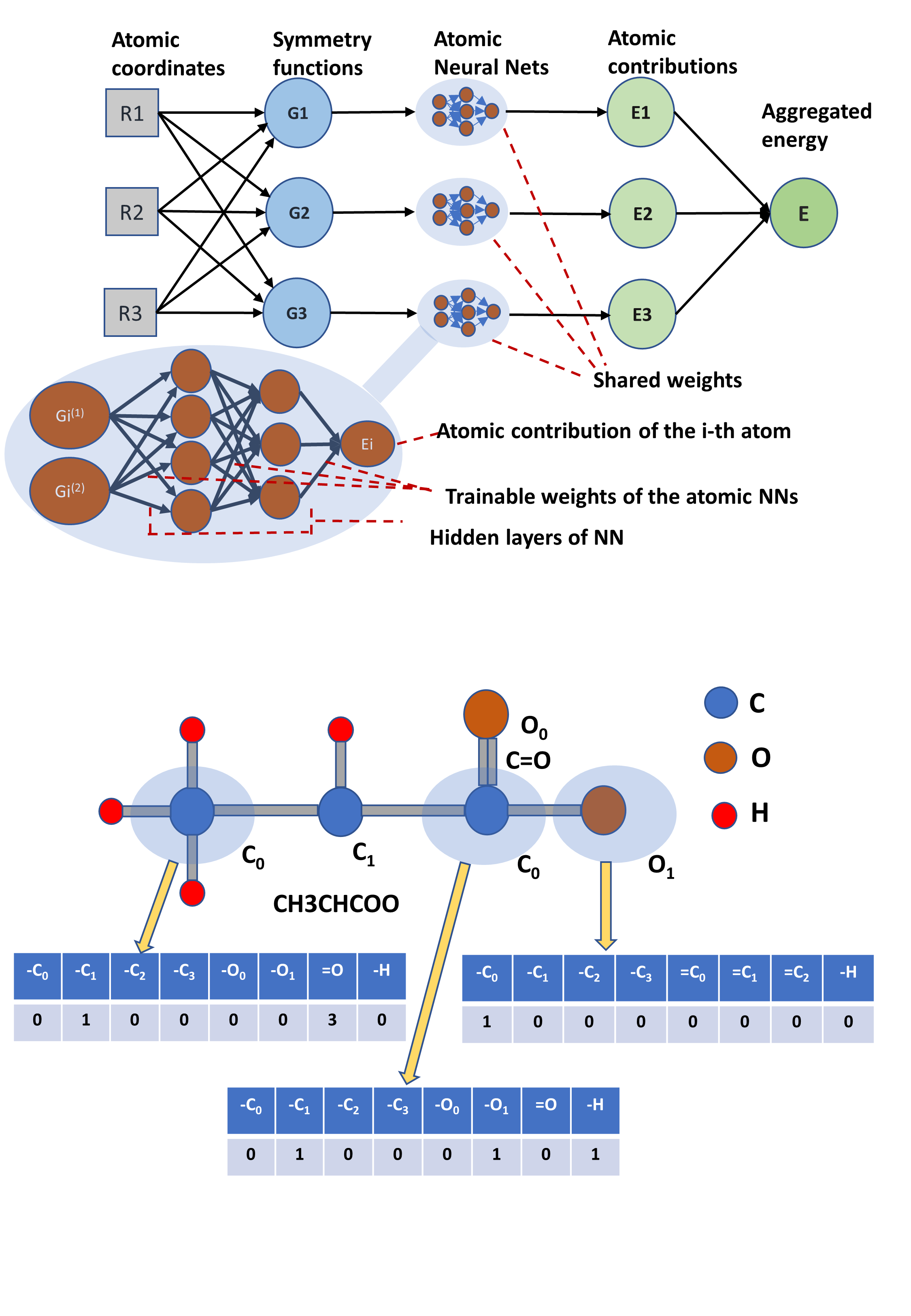}
\caption{The \textbf{top} figure shows the network for the atomic contribution method and the \textbf{bottom} figure presents our proposed non-coordinate based atomic fingerprints for this method. In the top figure, $Gi$ denotes the vector containing the symmetry function values for the i-th atom (calculated from the coordinates of all the atoms). For each atom, its corresponding G vector is fed to a neural network (NN). The structure of the atomic NN can be adjusted as shown in the bubble. The fingerprints can also be obtained directly from the SMILES notation of the molecule without the need of any atomic coordinates (bottom figure). Three sample fingerprint vectors for two carbon atoms and one oxygen atom are shown. The vectors contain the information about the number of different types of bonds for an atom. For example, the vector item $-C_1$ contains the number of single bonds the current atom has with carbon atoms that contain one free valence. The 5th to 7th positions of the fingerprint vectors have different meaning for carbon and oxygen atoms, e.g, in the 7th position, for carbon, $=O$ denotes how many oxygen the current carbon atom is bonded to by double bonds, whereas for oxygen, $=C_2$ encodes the number of carbon with two free valencies that the current oxygen atom is connected to by double bonds.} \label{fig:subnetimg}
\end{figure}
\subsection*{Additive Subnetwork Model}
The atomic fingerprint based additive subnetwork model is a size extensible model. The atomic fingerprint originally used~\cite{behlersymfun} for this model were the symmetry functions calculated from the atomic coordinates of all the atoms in the molecule. Two commonly used symmetry functions are: one that aggregates the pairwise distance information centered around each atom; the other that combines the angular distance information from a triplet of atoms (equations for these distance measures are given in the 18th page of the supporting information). Other symmetry functions can be devised, too. The model is shown in Figure~\ref{fig:subnetimg}. Fingerprints for each atom are fed to a small neural network. These subnetworks learn the energy contribution of the current atom to the total energy as a function of the fingerprints. Aggregated energy contributions from all the subnetworks yield the final energy. Subnetworks for all the atoms of an atom type share their weights. The weights can also be shared across the atom types. We have found the latter approach to work better for our case. The weight sharing ensures that the model is invariant to the ordering of the atoms. In contrast to other models, this one can only work with neural networks because it gives the flexibility of a hierarchical structure through the use of the back-propagation method to learn the network weights. In order to avoid the computation of reliable coordinates for the prediction set, we prefer the SMILES based fingerprints. We have developed such an atomic fingerprint, shown in Figure~\ref{fig:subnetimg}, that is similar to the flat molecular fingerprint described above, but is centered on an atom and encodes the connectivity information for that atom. We have found this model to work better for extrapolation compared to CM, BoB or flat fingerprints, but the errors were still quite large which warranted a further improvements to the model.

\subsection*{Proposed Multiple Filter based Additive Subnetwork Model}
To make the additive subnetwork based model generalize better to an unseen testing data, we propose to treat the shared weights of the atomic subnetworks as a filter in a convolutional neural network (CNN). This type of deep learning model is commonly used for image data where different sets of weights (called convolutional filter) scan through the image patches and learn to detect various basic image features such as edges or corners~\cite{cnnrep} which are combined in subsequent layers to detect higher level objects such as a face or a car or a digit~\cite{alexnet}. These filters are analogous to the local receptive field in biological visual systems~\cite{hubelwisel,recfield}. In CNN, the filters are usually 2D or 3D matrices of weights. A filter is placed on a patch of the image and a cross-correlation operation between the filter weights and the input plane pixels is performed - this is the output of that filter for that image patch. Then, the filter is moved to the next adjacent patch (which may or may not overlap with the previous patch). A key observation here is that there is not one but a number of filters that are used because each filter learns different features (through back-propagation)~\cite{lecundocrec}.

Moving to our problem and the atomic subnetworks, we can think of a subnetwork as a 
filter in CNN. Since all of the subnetworks share their weights, they can be considered as one filter scanning each of the atoms in the molecule one by one. Unlike CNN, however, in our case the subnetworks compute a non-linear function of its inputs instead of the cross-correlation, which makes sense since predicting adsorption energies is a regression problem and we want each subnetwork to learn the energy contribution of an atom. Also, unlike CNN, here we do not need multiple layers of filters as our learning objective is to find the individual atom contribution to the total energy. However, the aspect of CNN that can be incorporated in our networks and that can lend the additive subnetworks a better representational ability is to use multiple filters instead of one. Here, we should clarify that using multiple filters does not mean using separate filters for different atom types. Whether we use different shared weights for different atom types or not, by 'multiple filters' we are referring to completely separate sets of filters (in each set, there may be one filter if all atom types share the weights, or more than one if weights are shared only inside each atom type). In our proposed model, each of the separate set of filters would scan each atom of the molecule and the weighted sum of all the filtered values should yield the final output energy. 

\begin{figure}[H]
\centering
   \includegraphics[scale=0.68]{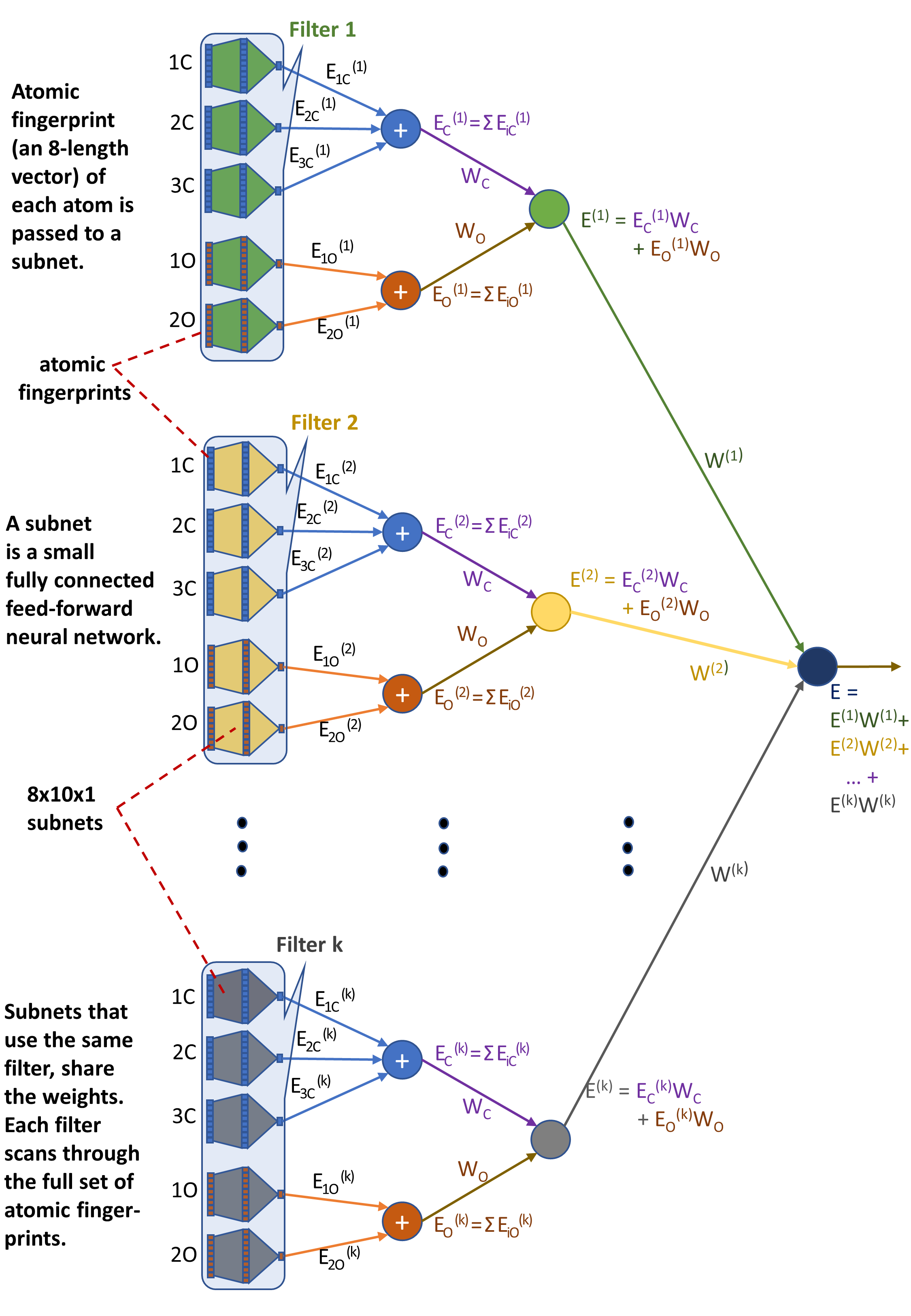}
\caption{Our proposed Model. A species with $3$ carbon atoms and $2$ oxygen atoms is passed through $k$ filters. Each filter is an $8$ by $10$ by $1$ neural network. For each filter, outputs of networks for each of the three C atoms is summed up, same is done for the two O atoms. The weighted sum of these two sums is the output for one filter. The final output is the weighted sum of all the filter outputs. The parameters of the network that are learned through back-propagation are: $W^{(i)}$s, $W_C$, $W_O$, and network weights for each filter.} \label{fig:repnet}
\end{figure}

The proposed model is shown in Figure~\ref{fig:repnet}. The atomic fingerprints are the $8$-length vectors from Figure~\ref{fig:subnetimg}. During the training of the network, at each iteration of the gradient descent, there is a forward pass that starts from the fingerprints at the left of the figure and moves to the right. The gradient of the error in the energy obtained at the right-most node is then back-propagated through the network which makes it learn the appropriate weights to fit the data. Nonlinearity in the computation comes from the nonlinear activation functions used in the hidden layers of the filters. The error function (that the gradient descent tries to optimize) for neural networks is non-convex~\cite{losssrf} - there can be many local minima. This means the output of a neural network is sensitive to the starting points of its weights - starting from different points in the hyperspace can result in different ending locations. Since the weights of each filter are randomly initialized~\cite{anninit,anninit2}, each of them is likely to end up with different weights than others even though all of them are fed the same set of atomic fingerprints, i.e, each of them learns different functions of their inputs, just like each CNN filter learns to detect different image features. Let us assume there are $K$ filters and the functions that they learn are denoted as $f^{(1)}, f^{(2)}, ..., f^{(K)}$ and the output of the $i$-th filter (after combining outputs of that filter for each atom in the species) is $E^{(i)}$. Then, the overall energy output of the networks, $E$ is
\begin{equation}
E = \sum_{i=1}^{K}E^{(i)}W^{(i)}
\end{equation}
where $W^{(i)}$ is the weight of the contribution for the $i$-th filter and
\begin{equation}
E^{(i)} = \sum_{a=1}^{T}E_{a}^{(i)}W_{a}
\end{equation}
where $T$ is the number of atom types in the species that have fingerprints (in our case, it is $2$, for C and O; since those two atom types can describe all the bonds inside the species, we have not included fingerprints for H), $E_{a}^{(i)}$ and $W_{a}$ are the summed contribution for all the atoms of atom type $a$ when passed through filter $i$, and the weight for that atom type which is shared across the filters, respectively. So,
\begin{equation}
E_{a}^{(i)} = \sum_{n=1}^{N_a}E_{a_n}^{(i)}
\end{equation}
where $E_{a_n}^{(i)}$ is the output when the $n$-th atom of atom type $a$ is passed through filter $i$. The last equation means the atomic contribution of all the atoms for an atom type for a filter are directly summed and not weighted (which can be treated like a constant, non-learnable weight of $1$). This ensures that a change in the relative ordering of the atoms (inside the set of atoms of an atom type) does not change the overall result. If the atomic fingerprint for the atom $a_n$ is $X_{a_n}$, then $E_{a_n}^{(i)}$ is a non-linear function of $X_{a_n}$:
\begin{equation}
E_{a_n}^{(i)} = f^{(i)}(X_{a_n})
\end{equation}
Here, the fingerprint is passed to the filter which, in our case, is a small fully connected feed-forward neural network (NN). The output of each layer in the NN is computed by multiplying the weight matrix between the current and the previous layer with the output vector of the previous layer and then passing the obtained vector to a non-linear activation function~\cite{compact,compact2,reluact}.

\section*{Results and Discussion}
\begin{table}[!htb]
\centering
\caption{Interpolation and extrapolation results. The methods used: Coulomb matrix, bag-of-bonds, flat molecular fingerprint, and additive atomic subnetwork model (See discussions for details). For the first seven rows of interpolation and the first four rows of extrapolation, for each of the methods, we ran the following ML models: ridge regression, LASSO, kernel ridge regression (KRR), support vector regression (SVR), Gaussian processes (GP). The rest of the rows used artificial neural networks (ANN). The first column denotes whether it is an interpolation or extrapolation. The second and third columns show the descriptor method and the ML model used, respectively. The fourth column contains the mean absolute error (MAE) of the predicted adsorption energies, and the fifth column presents the standard deviations of the absolute errors. For interpolation, data of $247$ C4 species were randomly permuted and $215$ were used for training and the rest for testing. The process was repeated $100$ times (data being permuted randomly each time) to obtain an unbiased estimate of the MAE. For extrapolation, data of $247$ C4 species were used for training and data of $29$ C2 and C3 species were used for testing.}
\label{interp}
\begin{adjustbox}{width=0.99\textwidth}
\begin{tabular}{|l|l|l|l|l|}
\toprule
 Prediction type & Method & ML model & MAE (eV) & SD of AE (eV) \\
\midrule
Interpolation & Coulomb matrix & GP & 0.230 & 0.218 \\ \hline
Interpolation & Bag-of-bonds & KRR & 0.139 & 0.136 \\ \hline
Interpolation & Bag-of-bonds & Ridge & 0.219 & 0.279 \\ \hline
Interpolation & ECFP & SVR & 0.165 & 0.179 \\ \hline
Interpolation & Flat molecular fingerprint (from SMILES) & SVR & 0.148 & 0.129 \\ \hline
Interpolation & Flat molecular fingerprint (from SMILES) & KRR & 0.141 & 0.122 \\ \hline
Interpolation & Flat molecular fingerprint (from SMILES) & Ridge & 0.196 & 0.166 \\ \hline
Interpolation & Additive atomic subnetwork & ANN & 0.398 & 0.202 \\ \hline
Interpolation & Proposed model (from coordinates, $1$ filter) & ANN & 0.347 & 0.259 \\ \hline
Interpolation & Proposed model (from coordinates, $4$ filters) & ANN & 0.309 & 0.231 \\ \hline
Interpolation & Proposed model (from SMILES, $1$ filter) & ANN & 0.190 & 0.164 \\ \hline
Interpolation & Proposed model (from SMILES, $6$ filters) & ANN & 0.142 & 0.120 \\ \hline
Extrapolation & Coulomb matrix & SVR & 2.392 & 1.015 \\ \hline
Extrapolation & Bag-of-bonds & KRR & 2.046 & 0.422 \\ \hline
Extrapolation & ECFP & SVR & 2.961 & 0.760 \\ \hline
Extrapolation & Flat molecular fingerprint (from SMILES) & KRR & 2.342 & 0.625 \\ \hline
Extrapolation & Additive atomic subnetwork & ANN & 0.441 & 0.214 \\ \hline
Extrapolation & Proposed model (from coordinates, $1$ filter) & ANN & 0.324 & 0.212 \\ \hline
Extrapolation & Proposed model (from coordinates, $4$ filters) & ANN & 0.282 & 0.190 \\ \hline
Extrapolation & Proposed model (from SMILES, $1$ filter) & ANN & 0.434 & 0.314 \\ \hline
Extrapolation & Proposed model (from SMILES, $5$ filters) & ANN & 0.227 & 0.143 \\
\bottomrule
\end{tabular}
\end{adjustbox}
\end{table}

In our investigations, we have used Coulomb matrix, bag-of-bonds, flat molecular fingerprints (non-coordinate based, calculated from the SMILES), and additive atomic subnetwork models for both interpolation and extrapolation. Key results are shown in Table~\ref{interp}. The supporting information contains full tables of all results. Here, the table for interpolation shows that non-coordinate based molecular fingerprints with kernel based ML models perform as good as coordinate based descriptors with the same ML models. The additive atomic subnetwork based on SMILES with multiple filter also worked well. For this model, we also used a coordinate based atomic fingerprint of length $5$ - aggregated pairwise distance measures from an atom to each of the four atom types involved (carbon, hydrogen, oxygen and top two layers of metal catalyst surface), plus the triplet distance measure. 

For extrapolation, both the Coulomb matrix and bag-of-bonds performed poorly. This is not unexpected since these methods are not size extensible and require padded zeros to make them work for different sized molecules. From this point of view, the flat molecular fingerprint comes as an attractive alternative as it is a constant sized descriptor (size of the molecule does not effect the size of the vector; no zero padding is required). But our results show that it performs no better than CM or BoB for extrapolation. However, the method that we found to be most promising was the additive atomic subnetwork. Since this method adds up the atomic contributions to the total energy, it is naturally size extensible. The initial predictive error obtained using this method (with the length $5$ fingerprint discussed above) was approximately $0.4$ eV. As a neural network ends up in a different location of its parameter hyperspace on different runs (because of randomly initialized parameters), we ran the model multiple times and our final result was an ensemble of these runs - for each target species, its predicted adsorption energy was the mean of its predicted values of all the runs. This yielded an extrapolation error of approximately $0.32$ eV. The ensemble method was used in all of our following models.

\begin{figure}[!htb]
\centering
   \includegraphics[scale=0.9]{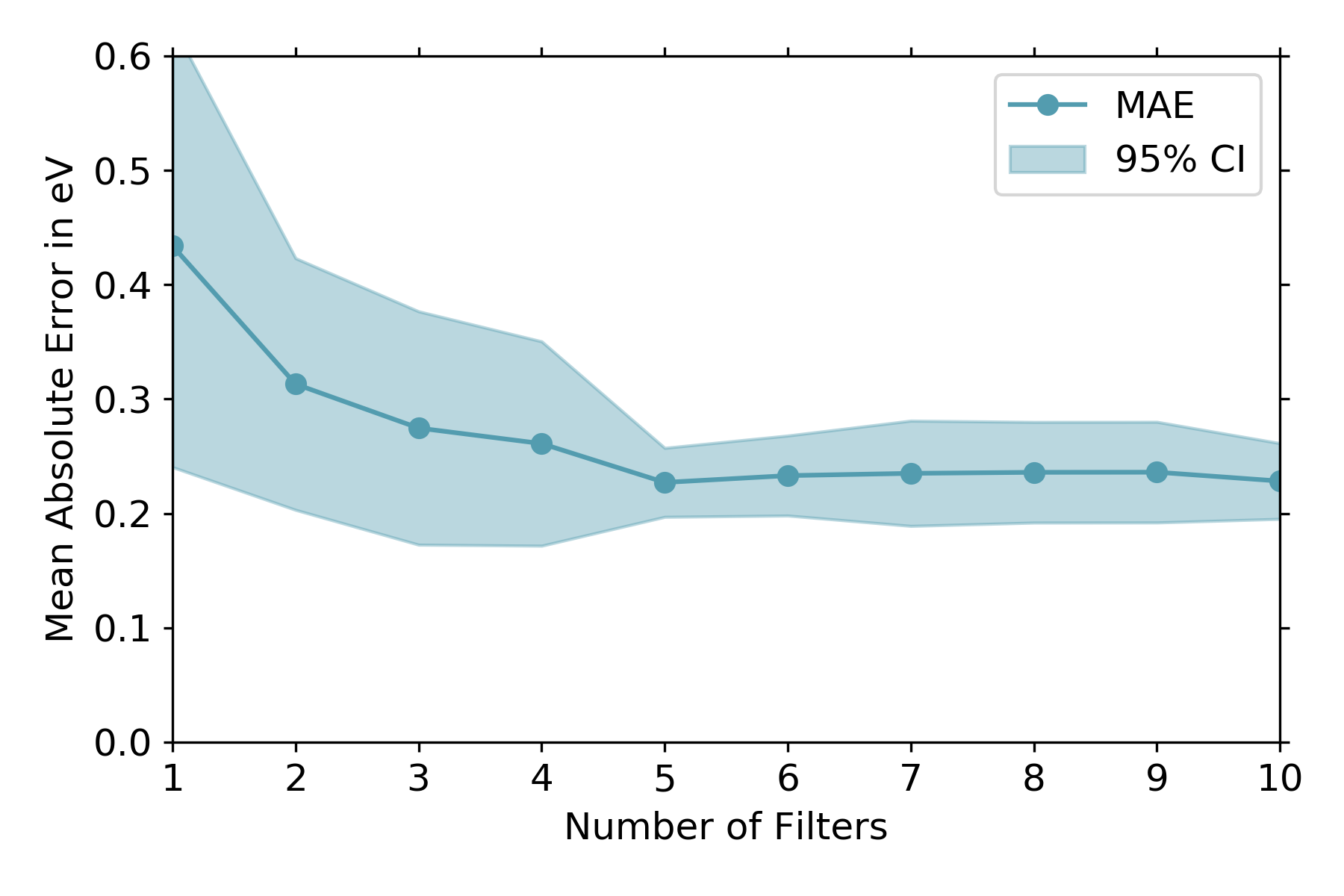}
\caption{Extrapolation errors decreased sharply with the use of more filters. At one point, however, it reaches a state where adding more filters does not make any significant improvement. The model used here is our proposed model shown in Figure~\ref{fig:repnet} and the atomic fingerprints were the ones shown in Figure~\ref{fig:subnetimg}. For each filter count, the predicted energies were obtained from the ensemble of $10$ runs of the proposed model. The process is repeated $100$ times for each filter count and the mean of the $100$ MAEs along with the confidence interval obtained from the standard deviation of the MAEs are shown in this figure.} \label{fig:errvsfilt}
\end{figure}

The next step was to replace the coordinate based atomic fingerprints with SMILES based ones (Figure~\ref{fig:subnetimg}). However, only replacing the atomic fingerprints in the additive subnetwork model actually increased the predictive errors to over $0.4$ eV. We improved this model by using multiple filters as discussed before (Figure~\ref{fig:repnet}). The predictive errors went down significantly as more filters were used. The rate of improvement, however, gradually subsided and after incorporating a certain number of filters the predictive errors change very little, as can be seen in Figure~\ref{fig:errvsfilt}. We used this multi-filter approach with the coordinate based atomic fingerprints as well, and the extrapolation error there went down to $0.28$ eV from over $0.32$ eV. 

\begin{figure}[!thb]
\centering
   \includegraphics[scale=0.65]{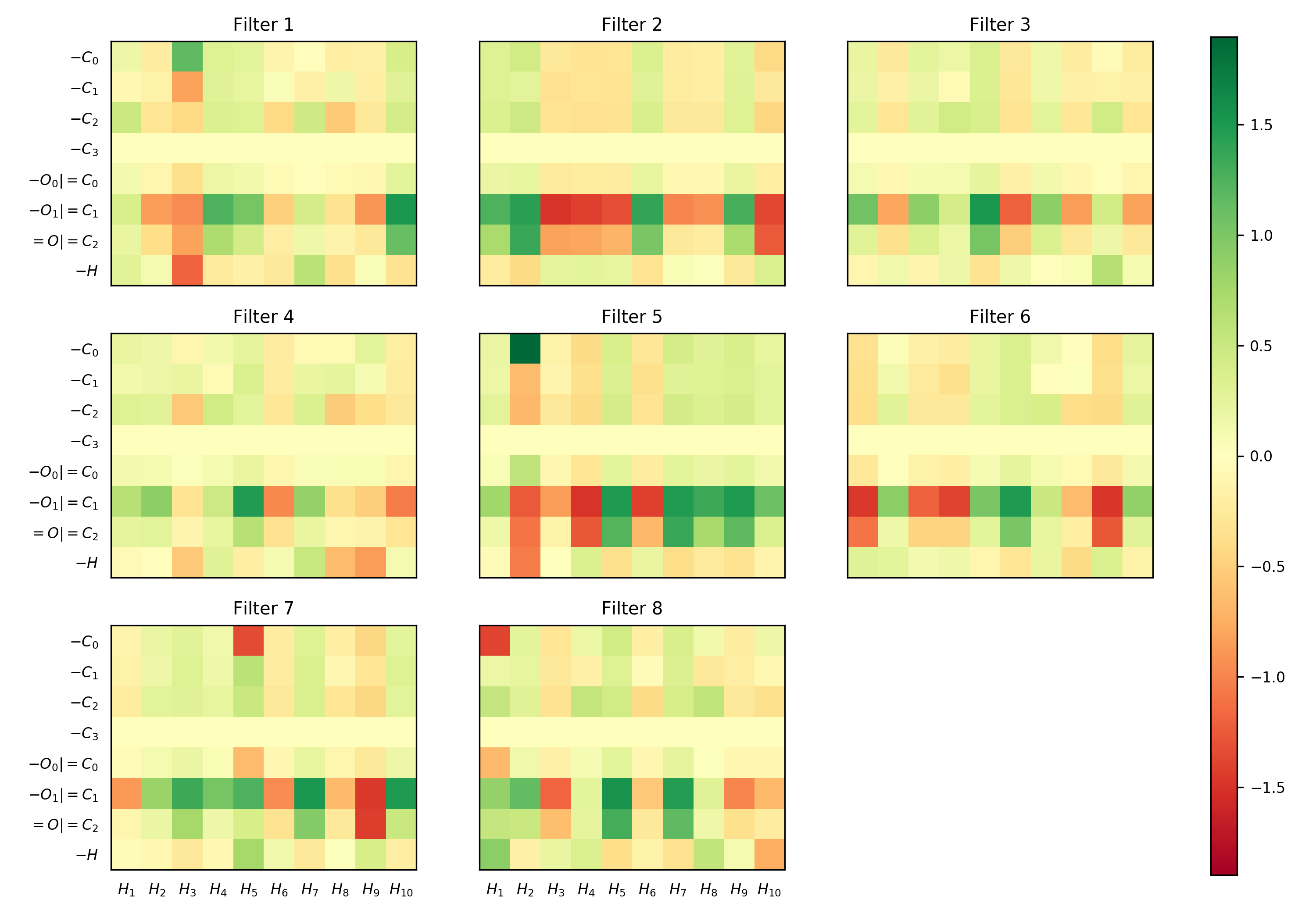}
\caption{Weight matrices for the $8$ filters learned by running our proposed model. Each of the eight $8$-by-$10$ matrices contains the values of the weights that connect each of the $8$ atomic fingerprint values (which, according to Figure~\ref{fig:subnetimg} are the number of different bond types an atom is connected to) to the $10$ hidden units denoted as $H_1$ to $H_{10}$ (the left half of each subnetwork or filter shown in Figure~\ref{fig:repnet}).} \label{fig:wtfilt}
\end{figure}

Here, we should note that, the number of filters is essentially a hyperparameter to our model and needs to be tuned for specific problems. Tuning of hyperparameters for machine learning models is typically done by setting aside a portion of the training set as validation set and choosing the hyperparameters for which the model performs best on the validation set. The chosen model is then used to run on the testing set. We also used this approach. This works well for interpolation problems where the training (which includes the validation set) and testing sets reside in the same region of the parameter space. But in case of extrapolation, this might not work. Through our investigations, we have seen that for extrapolation, the error on the validation set (which is part of the training set, containing C4 species) went to an approximate minimum value when $6$ filters were used and then remained more or less constant. But the extrapolation error on the testing set (containing C2 and C3 species), after the network was trained with different numbers of filters, reached minimum with $5$ filters. This can occur for other 'pure extrapolation' settings where a low validation error does not always correspond to a low test error. In this case, if a small amount of data from the test space (which, in our case, were C2 and C3 species) can be obtained, that can be included in the validation set to tune the hyperparameter more effectively. The problem setting, however, would no longer be a 'pure extrapolation' as small amount of data points from the test space are included during the training phase.

Figure~\ref{fig:wtfilt} shows the values of the learned weights between the input layer and the hidden layer for each filter when $8$ filters were used. For each matrix, a row corresponds to the $10$ weights going out of one fingerprint value (see Figure~\ref{fig:subnetimg}). A column corresponds to the $8$ values going into a hidden layer unit. There are three key observations here. First, each filer learns a different function of its inputs. Second, the sixth and seventh row contain weights with high absolute values. This is because the weights were shared across the atom types, i.e, fingerprints from carbon and oxygen atoms were fed to the same filters; and according to Figure~\ref{fig:subnetimg}, some of the entries at the same location of the fingerprint vector for carbon and oxygen carry a different meaning. According to this, the fifth, sixth, and seventh entries have to encode more information and hence the network learned higher valued weights for some of those. And finally, the fourth row for each of the filters learned close-to-zero weights. The fourth entry in the atomic fingerprint is the count of the number of carbon atoms that the current atom is bonded to where the carbon atom has three free valencies. In our training set, there was no such species. So, the network did not learn any significant value for those weights. This also signifies that if any species in the target set contains such a structure, its prediction would be inaccurate. Indeed, we found that two species, $CH_2C$ and $CH_3C$ (not included in the $29$ species used as our target set), both containing this type of structure, had very high prediction errors (around $1$ eV). This observation signifies a fundamental limitation in machine learning based models - the predictions can be at most as good as the data that is fed to train the model.

\section*{Conclusion}

In this paper, we have performed a detail investigation on a predictive model for both interpolation and extrapolation of adsorption energies of hydrocarbon species on a Pt(111) catalyst surface. We have compared the effectiveness of different fingerprints and ML models. For interpolation, our results indicate that a simple SMILE based fingerprint calculated from nearest neighbors with kernel based ML models performs very well for interpolation of adsorption energies with an MAE of $0.14$ eV. However, when predicting adsorption energies of species of different size from that of the training set (extrapolation), only an additive atomic contribution based model works reasonably well. To improve upon this method, we have developed a multi-filter based weighted additive model that combines the established additive model with the concept of filters from a convolutional neural network. Our findings show that this approach is highly generalizable compared to other models and leads for extrapolation of adsorption energies to an MAE of $0.23$ eV for a quite small dataset typical in computational catalysis. The proposed model also worked well when applied to interpolation with no statistically significant difference with the best models. The model has the potential to be applicable in other problems if the hyper-parameters of the model are adjusted according to the task. 

\bibliography{Extrapolation}

\begin{thebibliography}{10}

\bibitem{structdep}
J.~K. N\o{}rskov, F.~Studt, F.~Abild-Pedersen, T.~Bligaard, {\it Fundamental
  Concepts in Heterogeneous Catalysis\/} (John Wiley and Sons, Hoboken, New
  Jersey, 2014), chap.~2, pp. 17--19.

\bibitem{compcatal}
C.~Bo, F.~Maseras, N.~L{\'o}pez, {\it Nat. Catal.\/} {\bf 1}, 809 (2018).

\bibitem{lscl2}
J.~K. N{\o}rskov, F.~Abild-Pedersen, F.~Studt, T.~Bligaard, {\it Proc. Natl.
  Acad. Sci. U.S.A.\/} {\bf 108}, 937 (2011).

\bibitem{lscl1}
M.~Busch, M.~D. Wodrich, C.~Corminboeuf, {\it Chem. Sci.\/} {\bf 6}, 6754
  (2015).

\bibitem{linearscale}
F.~Abild-Pedersen, {\it et~al.\/}, {\it Phys. Rev. Lett.\/} {\bf 99}, 016105
  (2007).

\bibitem{theohet}
J.~P. Greeley, {\it Annu. Rev. Chem. Biomol.\/} {\bf 7}, 605 (2016).

\bibitem{difdom}
S.~Ben-David, {\it et~al.\/}, {\it Mach. Learn.\/} {\bf 79}, 151 (2010).

\bibitem{hdoprop}
J.~Lu, S.~Behtash, M.~Faheem, A.~Heyden, {\it J. Catal.\/} {\bf 305}, 56
  (2013).

\bibitem{gennn}
J.~Behler, M.~Parrinello, {\it Phys. Rev. Lett.\/} {\bf 98}, 146401 (2007).

\bibitem{dftml}
F.~Pereira, {\it et~al.\/}, {\it J. Chem. Inf. Model.\/} {\bf 57}, 11 (2017).
  PMID: 28033004.

\bibitem{selfpap1}
A.~J. Chowdhury, {\it et~al.\/}, {\it J. Phys. Chem. C\/} {\bf 122}, 28142
  (2018).

\bibitem{cmatrix}
M.~Rupp, A.~Tkatchenko, K.-R. M\"uller, O.~A. von Lilienfeld, {\it Phys. Rev.
  Lett.\/} {\bf 108}, 058301 (2012).

\bibitem{bob}
K.~Hansen, {\it et~al.\/}, {\it J. Phys. Chem. Lett.\/} {\bf 6}, 2326 (2015).

\bibitem{behlersymfun}
J.~Behler, {\it J. Chem. Phys.\/} {\bf 134}, 074106 (2011).

\bibitem{behlerref2}
T.~Morawietz, J.~Behler, {\it J. Phys. Chem. A\/} {\bf 117}, 7356 (2013). PMID:
  23557541.

\bibitem{behlerref3}
J.~Behler, {\it J. Chem. Phys.\/} {\bf 145}, 170901 (2016).

\bibitem{norssubnet}
Z.~W. Ulissi, {\it et~al.\/}, {\it ACS Catal.\/} {\bf 7}, 6600 (2017).

\bibitem{extconfp}
D.~Rogers, M.~Hahn, {\it J. Chem. Inf. Model.\/} {\bf 50}, 742 (2010). PMID:
  20426451.

\bibitem{chemoinf}
Y.~C. Lo, S.~E. Rensi, W.~Torng, R.~B. Altman, {\it Drug Discov. Today\/} {\bf
  23}, 1538 (2018).

\bibitem{invmol}
B.~Sanchez-Lengeling, A.~Aspuru-Guzik, {\it Science\/} {\bf 361}, 360 (2018).

\bibitem{molbench}
Z.~Wu, {\it et~al.\/}, {\it Chem. Sci.\/} {\bf 9}, 513 (2018).

\bibitem{graphconv}
S.~Kearnes, K.~McCloskey, M.~Berndl, V.~Pande, P.~Riley, {\it J. Comput. Aid.
  Mol. Des.\/} {\bf 30}, 595 (2016).

\bibitem{convfp}
D.~Duvenaud, {\it et~al.\/}, {\it Proceedings of the 28th International
  Conference on Neural Information Processing Systems - Volume 2\/}, NIPS'15
  (MIT Press, Cambridge, MA, USA, 2015), pp. 2224--2232.

\bibitem{krrml}
M.~Rupp, R.~Ramakrishnan, O.~A. von Lilienfeld, {\it J. Phys. Chem. Lett.\/}
  {\bf 6}, 3309 (2015).

\bibitem{rnnmolfp}
M.~H.~S. Segler, T.~Kogej, C.~Tyrchan, M.~P. Waller, {\it ACS Central Sci.\/}
  {\bf 4}, 120 (2018).

\bibitem{3dcnnbio}
W.~Torng, R.~B. Altman, {\it BMC Bioinform.\/} {\bf 18}, 302 (2017).

\bibitem{lilienconstdesc}
C.~R. Collins, G.~J. Gordon, O.~A. von Lilienfeld, D.~J. Yaron, {\it J. Chem.
  Phys.\/} {\bf 148}, 241718 (2018).

\bibitem{cnnrep}
Y.~LeCun, P.~Haffner, L.~Bottou, Y.~Bengio, {\it Shape, Contour and Grouping in
  Computer Vision\/} (Springer-Verlag, London, UK, UK, 1999), pp. 319--345.

\bibitem{alexnet}
A.~Krizhevsky, I.~Sutskever, G.~E. Hinton, {\it Commun. ACM\/} {\bf 60}, 84
  (2017).

\bibitem{hubelwisel}
D.~H. Hubel, T.~N. Wiesel, {\it J. Physiol.\/} {\bf 148}, 574 (1959).

\bibitem{recfield}
D.~L. Ringach, {\it J. Physiol.\/} {\bf 558}, 717 (2004).

\bibitem{lecundocrec}
Y.~{Lecun}, L.~{Bottou}, Y.~{Bengio}, P.~{Haffner}, {\it Proc. IEEE\/} {\bf
  86}, 2278 (1998).

\bibitem{losssrf}
A.~Choromanska, M.~Henaff, M.~Mathieu, G.~B. Arous, Y.~LeCun, {\it Proceedings
  of the Eighteenth International Conference on Artificial Intelligence and
  Statistics\/}, G.~Lebanon, S.~V.~N. Vishwanathan, eds. (PMLR, San Diego,
  California, USA, 2015), vol.~38 of {\it Proceedings of Machine Learning
  Research\/}, pp. 192--204.

\bibitem{anninit}
I.~Sutskever, J.~Martens, G.~Dahl, G.~Hinton, {\it Proceedings of the 30th
  International Conference on Machine Learning\/}, S.~Dasgupta, D.~McAllester,
  eds. (PMLR, Atlanta, Georgia, USA, 2013), vol.~28 of {\it Proceedings of
  Machine Learning Research\/}, pp. 1139--1147.

\bibitem{anninit2}
B.~Hanin, D.~Rolnick, {\it Advances in Neural Information Processing Systems
  31\/}, S.~Bengio, {\it et~al.\/}, eds. (Curran Associates, Inc., 2018), pp.
  571--581.

\bibitem{compact}
A.~Vehbi~Olgac, B.~Karlik, {\it International Journal of Artificial
  Intelligence And Expert Systems\/} {\bf 1}, 111 (2011).

\bibitem{compact2}
T.~G. Tan, J.~Teo, P.~Anthony, {\it Artif. Intell. Rev.\/} {\bf 41}, 1 (2014).

\bibitem{reluact}
A.~L. Maas, A.~Y. Hannun, A.~Y. Ng, {\it Proc. icml\/} {\bf 30}, 3 (2013).

\end{thebibliography}

\bibliographystyle{Science}

\section*{Acknowledgments}
This work has been supported by the United States Department of Energy, Office of Basic Energy Sciences (DE-SC0007167) (Heyden) and by the National Science Foundation under Grant No. DMREF-1534260 (Terejanu). K.E.A. acknowledges financial support from the National Science Foundation under Grant No. OIA-1632824. Computational resources provided by XSEDE facilities located at San Diego Supercomputer Center (SDSC) and Texas advanced Computing Center (TACC) under grant number TG-CTS090100, U.S. Department of Energy facilities located at the National Energy Research Scientific Computing Center (NERSC) under Contract No. DE-AC02-05CH11231 and Pacific Northwest National Laboratory (Ringgold ID 130367, Grant Proposal 49246) and the High-Performance Computing clusters located at University of South Carolina are gratefully acknowledged.     
\section*{Supplementary materials}
Data for energy and SMILES for species.\\
Calculation method and descriptor values for CM, BoB, flat molecular fingerprints, atomic fingerprints.\\
Hyper-parameters for the proposed model.\\
Tables for complete results of interpolation and extrapolation.

\end{document}